\documentclass[preprint,tightenlines,superscriptaddress,showpacs,byrevtex]{revtex4}
\usepackage{epsfig}
\usepackage{multirow}

\usepackage{graphicx} 
\usepackage{dcolumn} 
\usepackage{rotating}
\usepackage{adjustbox}
\usepackage{threeparttable}
\usepackage[colorlinks,linkcolor=red,anchorcolor=green,citecolor=blue]{hyperref}
\usepackage{verbatim}
\usepackage{amsmath}

\usepackage{makecell}
\usepackage{pgffor}
\usepackage{xparse}
\usepackage{comment}
\usepackage{lineno}
\usepackage{mciteplus}
\usepackage[T1]{fontenc}

\def\ee{\ensuremath{e^+e^-}}
\def\mumu{\ensuremath{\mu^+\mu^-}}
\def\GeV{\ifmmode {\mathrm{\ Ge\kern -0.1em V}}\else
                   \textrm{Ge\kern -0.1em V}\fi}%
\def\MeV{\ifmmode {\mathrm{\ Me\kern -0.1em V}}\else
                   \textrm{Me\kern -0.1em V}\fi}%
\def\GeVcc{\ifmmode {\mathrm{\ Ge\kern -0.1em V}/c^2}\else
                   {\textrm{Ge\kern -0.1em V}/$c^2$}\fi}%
\def\MeVcc{\ifmmode {\mathrm{\ Me\kern -0.1em V}/c^2}\else
                   {\textrm{Me\kern -0.1em V}/$c^2$}\fi}%
\def\rts{\ensuremath{\sqrt{s}}}

\NewDocumentCommand{\Ds}{st0t1t+t-}{%
	\ensuremath{D_{s\IfBooleanT{#2}{0}\IfBooleanT{#3}{1}}^{%
			\IfBooleanT{#1}{*}%
			\IfBooleanTF{#3}{}{\IfBooleanF{#2}{
				\IfBooleanTF{#4}{%
					\IfBooleanTF{#5}{\pm}{+}%
				}{%
					\IfBooleanT{#5}{-}%
		}}}}%
		\IfBooleanT{#3}{(2460)^{%
			\IfBooleanTF{#4}{%
				\IfBooleanTF{#5}{\pm}{+}%
			}{%
				\IfBooleanT{#5}{-}%
		}}}%
		\IfBooleanT{#2}{(2317)^{%
			\IfBooleanTF{#4}{%
				\IfBooleanTF{#5}{\pm}{+}%
			}{%
				\IfBooleanT{#5}{-}%
		}}}%
	}%
}

\begin{document}

\title{\boldmath Measurement of the Born Cross Sections for $e^+e^-\to D_s^+ D_{s1}(2460)^- +c.c.$ and $e^+e^-\to D_s^{\ast +} D_{s1}(2460)^- +c.c.$}

\author{
M.~Ablikim$^{1}$, M.~N.~Achasov$^{10,d}$, P.~Adlarson$^{64}$, S. ~Ahmed$^{15}$, M.~Albrecht$^{4}$, A.~Amoroso$^{63A,63C}$, Q.~An$^{60,48}$, ~Anita$^{21}$, Y.~Bai$^{47}$, O.~Bakina$^{29}$
% ...
\\
\vspace{0.2cm}
(BESIII Collaboration)\\
\vspace{0.2cm} {\it
$^{1}$ Institute of High Energy Physics, Beijing 100049, People's Republic of China\\
% ...
$^{4}$ Bochum Ruhr-University, D-44780 Bochum, Germany\\
% ...
$^{10}$ G.I. Budker Institute of Nuclear Physics SB RAS (BINP), Novosibirsk 630090, Russia\\
% ...
$^{15}$ Helmholtz Institute Mainz, Johann-Joachim-Becher-Weg 45, D-55099 Mainz, Germany\\
% ...
$^{21}$ Indian Institute of Technology Madras, Chennai 600036, India\\
% ...
$^{29}$ Joint Institute for Nuclear Research, 141980 Dubna, Moscow region, Russia\\
% ...
$^{47}$ Southeast University, Nanjing 211100, People's Republic of China\\
$^{48}$ State Key Laboratory of Particle Detection and Electronics, Beijing 100049, Hefei 230026, People's Republic of China\\
% ...
$^{60}$ University of Science and Technology of China, Hefei 230026, People's Republic of China\\
% ...
$^{63}$ (A)University of Turin, I-10125, Turin, Italy; (B)University of Eastern Piedmont, I-15121, Alessandria, Italy; (C)INFN, I-10125, Turin, Italy\\
$^{64}$ Uppsala University, Box 516, SE-75120 Uppsala, Sweden\\
% ...
$^{d}$ Also at the Novosibirsk State University, Novosibirsk, 630090, Russia\\
% ...
}
}

\begin{abstract}

	The processes $e^+e^-\to D_s^+ D_{s1}(2460)^- +c.c.$ and $e^+e^-\to D_s^{\ast +} D_{s1}(2460)^- +c.c.$ are studied for the first time using data samples collected with the BESIII detector at the BEPCII collider. The Born cross sections of $e^+e^-\to D_s^+ D_{s1}(2460)^- +c.c.$ at nine center-of-mass energies between 4.467\,GeV and 4.600\,GeV and those of $e^+e^-\to D_s^{\ast +} D_{s1}(2460)^- +c.c.$ at ${\sqrt s}=$ 4.590\,GeV and 4.600\,GeV are measured.  No obvious charmonium or charmonium-like structure is seen in the measured cross sections.

\end{abstract}


\maketitle

\section{\boldmath INTRODUCTION}

The charmed-strange mesons, known as \Ds, are made up of $c\bar{s}$ or $\bar{c}s$ quarks. The \Ds1 meson was first observed in 2003 by the CLEO experiment via its decay into $D_s^{*+}\pi^0$~\cite{Ds1(2460)CLEO2003}. It was subsequently confirmed by the Belle~\cite{Ds1(2460)Belle2003} and BABAR~\cite{Ds1(2460)Babar2004} experiments. The experimental results favor a $J^P=1^+$ quantum number assignment for \Ds1 as a $P$-wave state. However, its measured mass $(2459.5\pm 0.6)\MeVcc$ is at least $70\MeVcc$ lower than the quark model predictions~\cite{potModel1985,potModel1995}, leading to an unexpectedly narrow width. It has also been proposed to be a good candidate for a $D^*K$ molecule state~\cite{D*Kmolecule2007,radioDecay2014,2460trans2016}, or a mixture of the $c\bar{s}$ and $D^*K$ state~\cite{mole2016}.

The \Ds1 can be produced in the processes $\ee\to\Ds+\Ds1- +c.c.$ and $\ee\to\Ds*+\Ds1- +c.c.$. Following the excitation behavior of $S$-wave production, Ref.~\cite{DsDs1theory} predicts
$\sigma\left[\ee\to\Ds*\Ds*0\right]$ and $\sigma\left[\ee\to\Ds\Ds1\right]\propto\sqrt{E_{\rm c.m.}-E_0}$, where $E_{\rm c.m.}$ is the center-of-mass (c.m.) energy and $E_0\approx 4.43$ GeV is the mass threshold of both channels.

Additionally, several charmonium-like $Y$ states with $J^{PC}=1^{--}$ lying above the open charm threshold have been discovered, such as the $Y(4260)$~\cite{Y42601,Y42602,Y42603}, $Y(4360)$~\cite{Y4360,Y4660}, and $Y(4660)$~\cite{Y4660}.
Measurements of these charmonium-like states decaying into a charmed-antistrange and anticharmed-strange meson pair provide crucial insight on their internal structure. The Belle~\cite{DsDsBelle}, BABAR~\cite{DsDsBaBar}, and CLEO~\cite{DsDsCLEO} experiments have measured the cross sections of $\ee\to D_s^{(*)}\bar{D}_s^{(*)}$ with low-lying charmed-strange mesons in the final states. Using an $e^+e^-$ collision data sample corresponding to 567~$\mathrm{pb}^{-1}$ collected at $\sqrt{s}= 4.600$\,GeV, the BESIII experiment has measured the cross section of $\ee\to\Ds+\bar{D}^{(*)0}K^-$, which includes significant contributions from events with the $D_{s1}(2536)^-$ and $D_{s2}^*(2573)^-$ charmed-strange mesons~\cite{DsDKBES}.
Using a data sample of 921.9 fb$^{-1}$ collected at $\sqrt{s}=10.52$, 10.58, and 10.867 GeV, Belle
measured the cross sections of $\ee \to \Ds+ D_{s1}(2536)^-$ and $\ee \to \Ds+ D_{s2}^*(2573)^-$ and observed the $Y(4626)$
with significances of 5.9$\sigma$ and 3.4$\sigma$, respectively, with systematic uncertainties included~\cite{sen1,sen2}.

In this paper, we report the first measurement of the Born cross sections for $\ee\to\Ds+\Ds1- +c.c.$ and $\ee\to\Ds*+\Ds1- +c.c.$, and the search for possible vector charmonium-like states. Throughout the paper, charged-conjugate modes are always implied.

\section{\boldmath DETECTOR, DATA SAMPLES AND MONTE CARLO SIMULATIONS}

BESIII~\cite{besiii1} and BEPCII are major upgrades of the BESII detector~\cite{besiii2} and the BEPC accelerator. The cylindrical core of the BESIII detector consists of a helium-based multilayer drift chamber (MDC), a plastic scintillator time-of-flight system (TOF), and a CsI(Tl) electromagnetic calorimeter (EMC), which are all enclosed in a superconducting solenoidal magnet providing a 1.0 T  magnetic field. The solenoid is supported by an octagonal flux-return yoke with resistive plate counter muon identifier modules interleaved with steel. The acceptance of charged particles and photons is 93\% over 4$\pi$ solid angle.
The charged particle momentum resolution at 1 GeV/c is 0.5\%, and the energy loss ($\mathrm{d}E/\mathrm{d}x$) resolution is 6\% for the electrons from Bhabha scattering. The EMC photon energy resolution is 2.5\% (5\%) at 1 GeV in the barrel (end cap) region. The time resolution of the TOF barrel (end cap) is 68 ps (110 ps).
Our particle identification (PID) methods combine the TOF information with the ${\rm d}E/{\rm d}x$ measured in the MDC to calculate the probability Prob($h$), $h=\pi,K$, for a track to be a pion or a kaon.

In this paper, the Born cross sections of the processes
$\ee\to\Ds+\Ds1-$ and $\ee\to\Ds*+\Ds1-$ are measured for the first time at nine energy points between 4.467 and 4.600~GeV, and at 4.590 and 4.600~GeV, respectively.
Table~\ref{tab:1} lists the data samples used in this analysis and their integrated luminosities.
The c.m.\ energies are measured using the process $\ee\to\mumu$ with an uncertainty of 0.8\,MeV~\cite{XYZEcm}.
The integrated luminosities are measured with an uncertainty of 1.0\%
using large-angle Bhabha scattering events~\cite{XYZLum, RscanLum}.

The {\sc geant4}-based~\cite{GEANT} Monte Carlo (MC) simulation framework {\sc boost}~\cite{boost}, which consists of event generators and the description of the detector geometry and response, is used to produce large simulated event samples. These are used to optimize the event selection criteria, determine the detection efficiency, evaluate the initial state radiation (ISR) correction factor ($1+\delta$), and estimate background contributions.
The simulation includes the beam energy spread and ISR modeled with {\sc kkmc}~\cite{KKMC,KKMC2,KKMC3} and {\sc besevtgen}~\cite{besevtgen,besevtgen2}.
The final state radiation (FSR) effects are simulated by the {\sc photos}~\cite{photos} package.
For each energy point, we generate MC samples of the signal processes $\ee\to\Ds+\Ds1-$ and $\ee\to\Ds*+\Ds1-$ with a uniform distribution in phase space (PHSP).

The signal process $\ee\to\Ds+\Ds1-$ is simulated with $\Ds+$ decaying into
$K^+K^-\pi^+$, and the $\Ds1-$ decaying into all possible final states.
The signal process $\ee\to\Ds*+\Ds1-$ is simulated with $\Ds*+$ decaying into $\gamma\Ds+$ and the $\Ds1-$ decaying into all possible final states.
A $P$-wave model and a Dalitz plot decay model~\cite{dpgen} are used to simulate $\Ds*+\to\gamma\Ds+$ and
$\Ds+\to K^+K^-\pi^+$, respectively.

Two generic MC simulated samples at $\rts=$ 4.575 GeV and 4.600 GeV, equivalent to the respective integrated luminosity of each data set, are produced to investigate potential
peaking background channels. Known processes and decay modes are generated by {\sc besevtgen} with cross sections and branching fractions obtained from the Particle Data Group (PDG)~\cite{PDG2018}. The remaining unmeasured phenomena associated with charmonium decays or open charm processes are simulated with {\sc lundcharm}~\cite{besevtgen,lundcharm}, while continuum light hadronic events are produced with {\sc pythia}~\cite{pythia}.

\section{\boldmath COMMON SELECTION CRITERIA}
\label{selection}
The candidate events for $\ee\to\Ds+\Ds1-$ and $\ee\to\Ds*+\Ds1-$ are selected with a partial reconstruction method to obtain higher efficiencies.
The \Ds+ candidates are reconstructed via $D_s^+\to\phi\pi^+, \phi\to K^+K^-$ and $\Ds+\to \bar{K}^{*0}K^+$, $\bar{K}^{*0}\to K^-\pi^+$. The \Ds*+ candidates are reconstructed via $\Ds*+\to\gamma\Ds+$. The \Ds1- signals are identified with the mass recoiling against the reconstructed \Ds+ and \Ds*+.
There are three charged tracks in $D_s^+\to K^+K^-\pi^+$, and one additional photon candidate in $D_s^{*+}\to\gamma D_s^+$.

For each charged track candidate, the polar angle $\theta$ in the MDC with respect to the detector axis must satisfy $|\cos\theta|<$ 0.93, and the point of closest approach to the $\ee$ interaction point must be within $\pm$10 cm in the beam direction and within 1 cm in the plane perpendicular to the beam direction. Pion candidates are required to satisfy ${\rm Prob}(\pi)>{\rm Prob}(K)$ and ${\rm Prob}(\pi)>0.001$. Kaon candidates are required to satisfy ${\rm Prob}(K)>{\rm Prob}(\pi)$ and ${\rm Prob}(K)>0.001$.

The photon candidates are selected from showers in the EMC. The deposited energy in the EMC is required to be larger than 25 MeV in the barrel region ($\left|\cos\theta\right|<0.80$) or greater than 50 MeV in the endcap region ($0.86<\left|\cos\theta\right|<0.92$). To eliminate the showers produced by charged tracks, photon candidates must be separated by at least $20^\circ$ from the extrapolated position of all charged tracks in the EMC. The timing of the shower is required to be within 700 ns from the reconstructed event start time to suppress noise and energy deposits unrelated to the event.

The candidate events of both $\ee\to\Ds+\Ds1-$ and $\ee\to\Ds*+\Ds1-$ are required to contain at least two kaons and one pion. One additional photon candidate is required for $\ee\to\Ds*+\Ds1-$. All combinations of $K^+K^-\pi^+$ that pass the vertex fit are kept.
To select $D_s^+\to\phi\pi^+, \phi\to K^+K^-$ and $\Ds+\to \bar{K}^{*0}K^+$, $\bar{K}^{*0}\to K^-\pi^+$ sub-modes, the invariant masses of $K^+K^-$ and $K^-\pi^+$ are required to satisfy $|M(K^+K^-)-m_{\phi}|<15$\,MeV/$c^2$ and $|M(K^-\pi^+)-m_{\bar{K}^{*0}}|<84$ MeV/$c^2$, respectively, where $m_{\phi}$ ($m_{\bar{K}^{*0}}$) is the nominal mass of the $\phi$ ($\bar{K}^{*0}$) meson taken from the PDG~\cite{PDG2018}.

\section{\boldmath MEASUREMENT OF $\ee\to\Ds+\Ds1-$ }

To improve the resolution of the \Ds+ recoil mass, we define $M_{\Ds+}^{\rm rec}\equiv M_{K^+K^-\pi^+}^{\rm recoil}+M(K^+K^-\pi^+)-m_{\Ds+}$, where $M_{K^+K^-\pi^+}^{\rm recoil}=
\sqrt{\left(P_{\rm c.m.}-P_{K^+}-P_{K^-}-P_{\pi^+}\right)^2}$, $P_{\rm c.m.}$, $P_{K^+}$, $P_{K^-}$, and $P_{\pi^+}$ are the four-momenta of the initial $\ee$ system, the selected $K^+$, $K^-$, and $\pi^+$, respectively, $M(K^+K^-\pi^+)$ is the invariant mass of the $K^+K^-\pi^+$ system, and $m_{\Ds+}$ is the nominal mass of the \Ds+ meson~\cite{PDG2018}.

\begin{figure}[htbp!]
	\begin{center}
	\includegraphics[width=.4\textwidth]{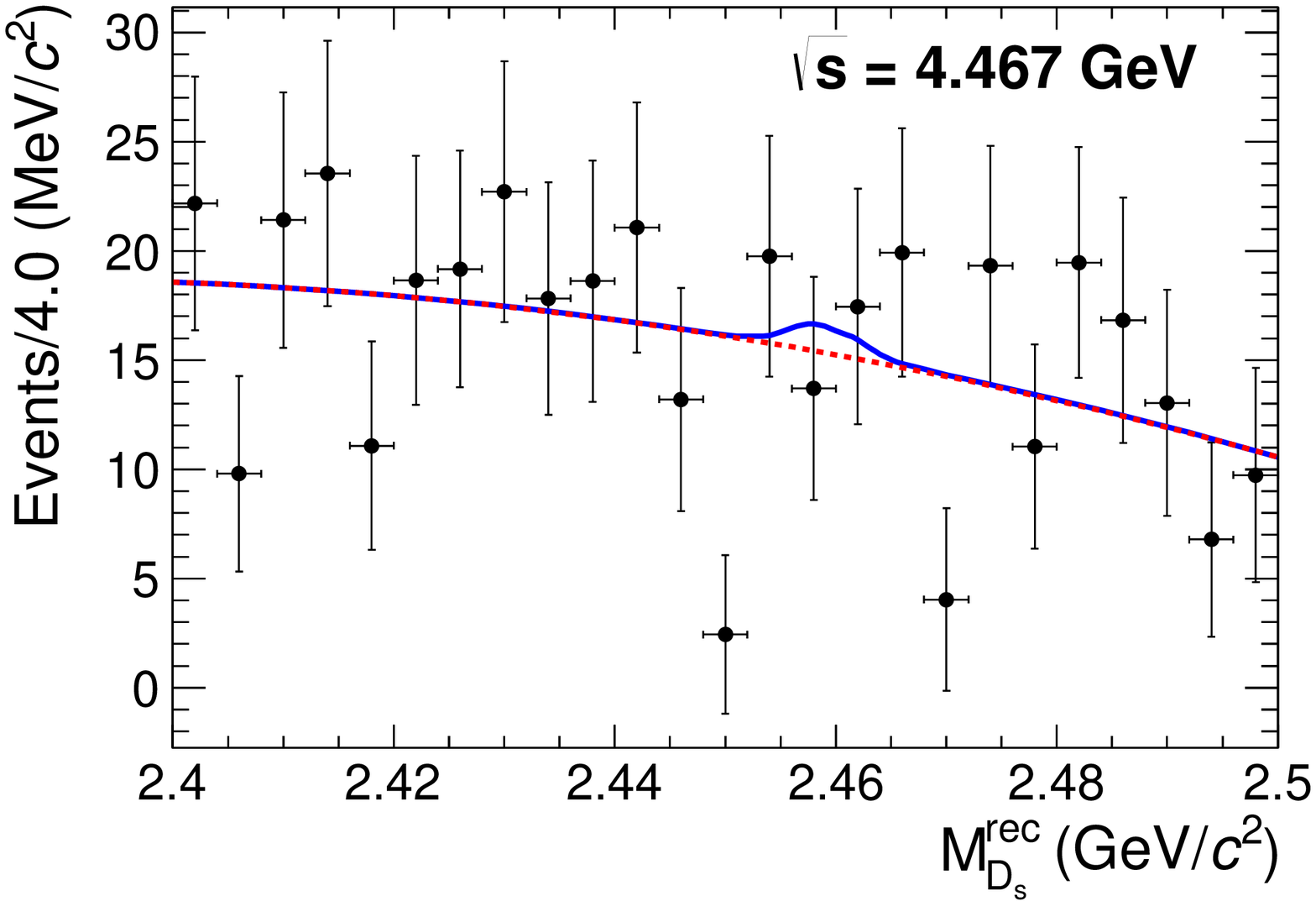}
	\includegraphics[width=.4\textwidth]{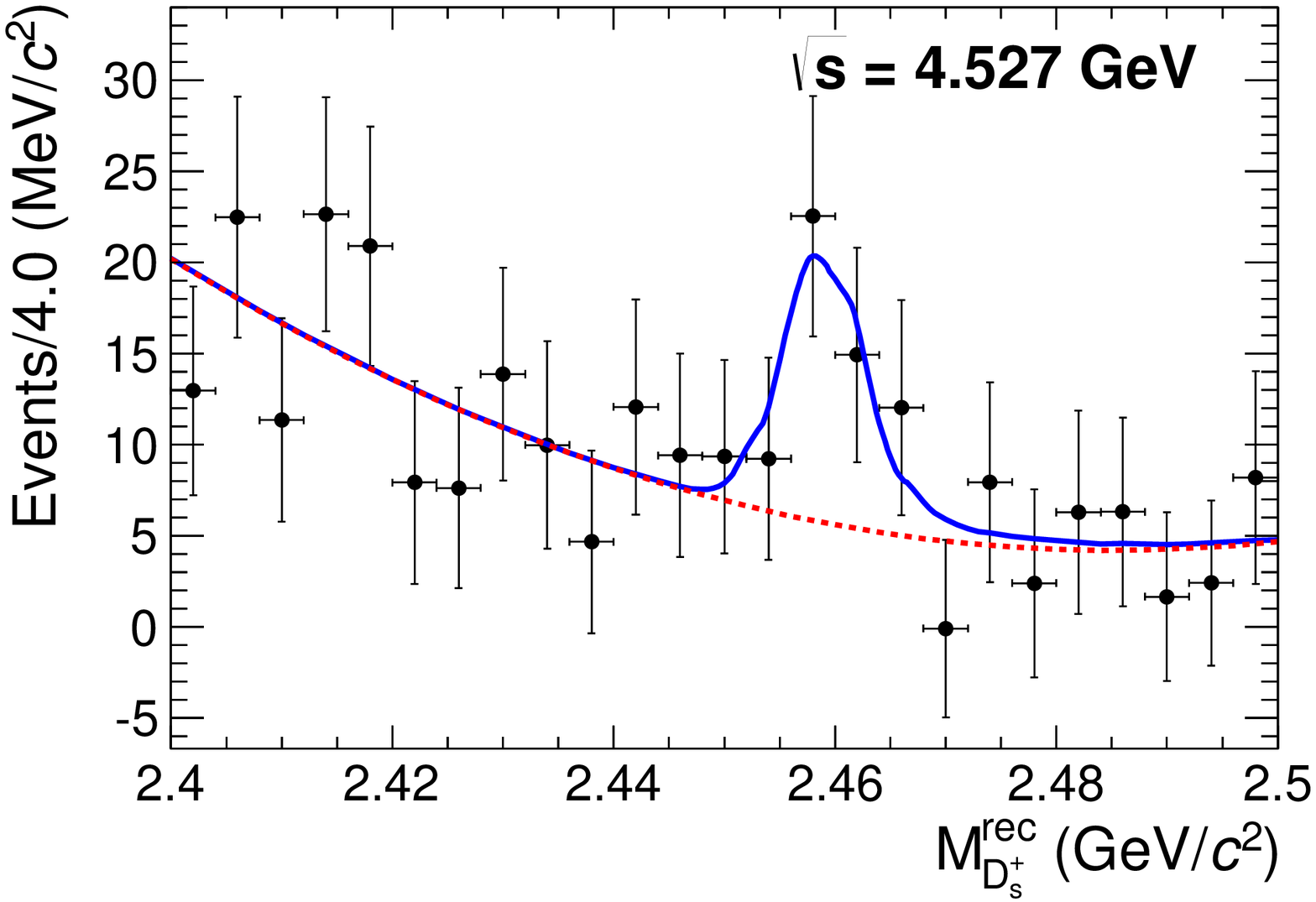}
	\includegraphics[width=.4\textwidth]{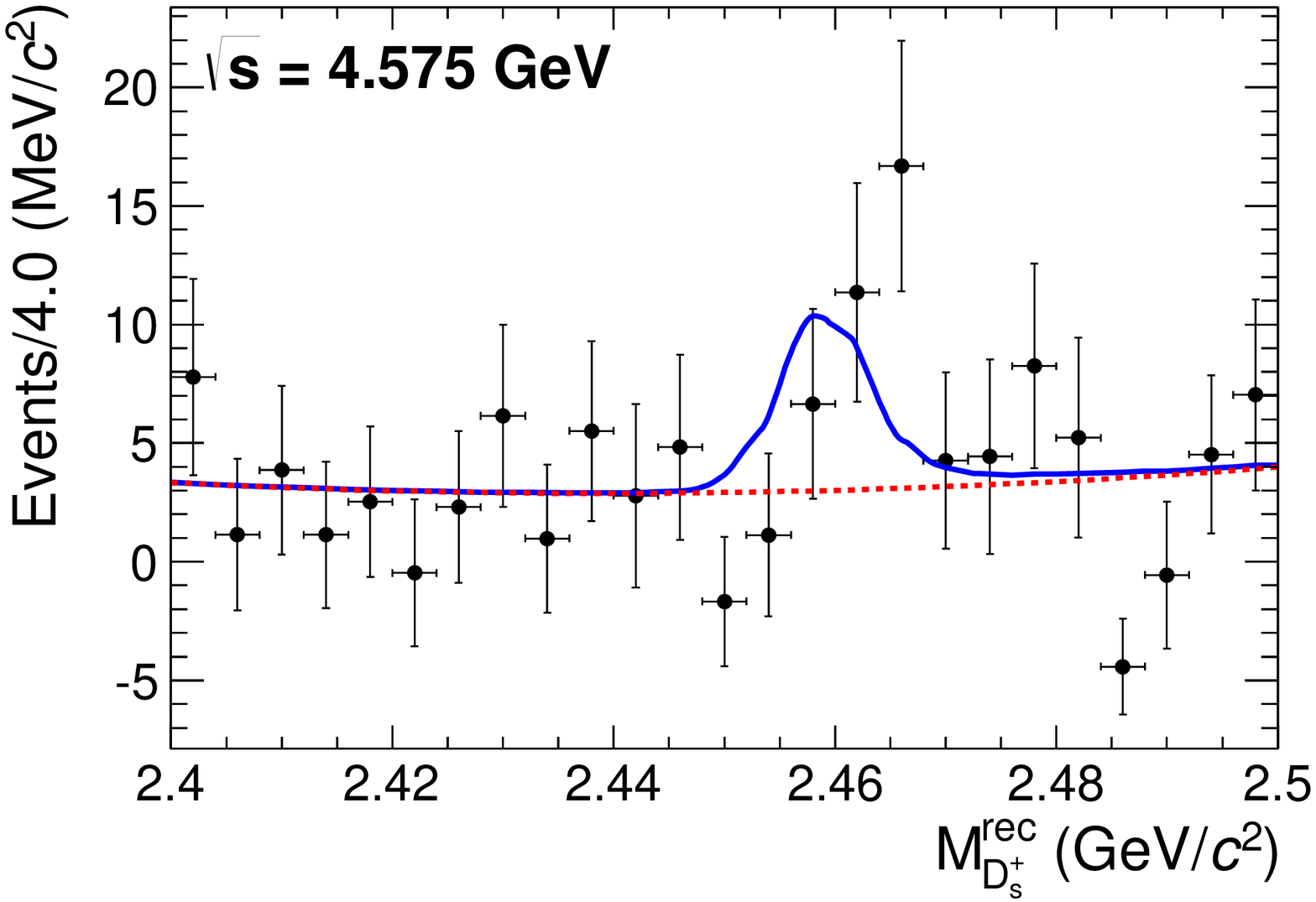}
	\includegraphics[width=.4\textwidth]{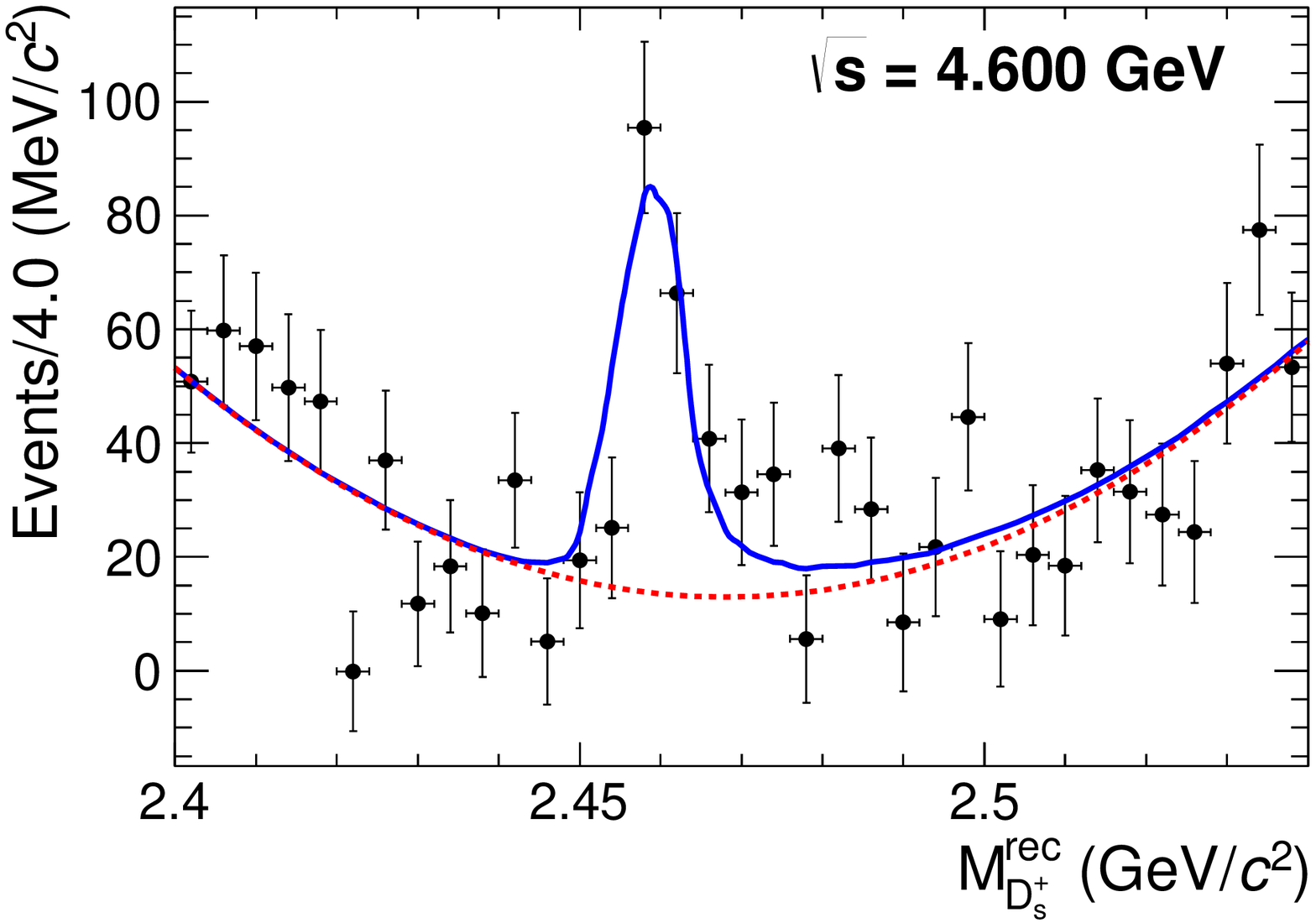}
	\caption{$M_{\Ds+}^{\rm rec}$ distributions at $\sqrt{s}=$ 4.467~GeV, 4.527~GeV, 4.575~GeV, and 4.600~GeV, respectively, obtained by extracting $D_s^+$ signal yields in the fit to the $M(K^+K^-\pi^+)$ distribution in each $M_{\Ds+}^{\rm rec}$ bin. The dots with error bars are data, the solid lines are the best fits, and the dashed lines are the fitted backgrounds. Clear \Ds1- signals are seen at $\sqrt{s}=$ 4.527~GeV, 4.575~GeV, and 4.600~GeV. The fitted results together with the signal significances are summarized in Table~\ref{tab:1}.}
	\label{fig:ds_sigma_data}
	\end{center}
\end{figure}

We separate the $M_{\Ds+}^{\rm rec}$ spectrum into 4.0\,MeV/$c^2$ wide bins.
We use 25 bins between 2.40 GeV/$c^2$ and 2.50 GeV/$c^2$ for the data samples taken at $\sqrt{s}=$ 4.467~GeV, 4.527~GeV, and 4.575~GeV, and 35 bins between 2.40 GeV/$c^2$ and 2.54 GeV/$c^2$ for the data sample at $\sqrt{s}=$ 4.600 GeV.
An unbinned maximum likelihood fit is performed to the $M(K^+K^-\pi^+)$ distribution for events in each $M_{\Ds+}^{\rm rec}$ bin.
The signal distribution is modeled by a Gaussian function, the parameters of which are fixed to those obtained from the fit to the original integrated $M(K^+K^-\pi^+)$ spectrum. The background shape is described by a first-order polynomial function. The obtained $M_{\Ds+}^{\rm rec}$ distributions, based on these fitted $D_s^+$ signal yields, are shown in Fig.~\ref{fig:ds_sigma_data} for four different energy points.
Detailed studies of the generic MC samples~\cite{topo} indicate that there are no peaking backgrounds in the $\Ds1-$ signal region. In the lower mass region the dominant backgrounds are from the process $e^+e^- \to D_s^{*+}D_s^{*-}$, while in the higher mass region the backgrounds are from processes with final states $D_s^+\bar{D}^{(*)0}K^-$, $D_s^+D^{(*)-}\bar{K}^0$, etc.

We fit these $M_{\Ds+}^{\rm rec}$ distributions to determine the signal yield of \Ds1-. The signal distribution is modeled by a MC-derived signal shape, while the background is described by a second-order polynomial. The fit results are shown in Fig.~\ref{fig:ds_sigma_data} and summarized in Table~\ref{tab:1}.
The significances of the \Ds1- signals are determined from the changes in the log-likelihood values with and without inclusion of a \Ds1- signal in the fit, taking the change of the number of degrees of freedom into account.
We obtain significances larger than 3$\sigma$ at $\sqrt{s}=$ 4.527~GeV, 4.575~GeV, and 4.600\,GeV. No significant \Ds1- signal is observed in the data sample at $\sqrt{s}=$ 4.467~GeV.

\begin{figure}[htbp!]
	\begin{center}
	\includegraphics[width=.28\textwidth]{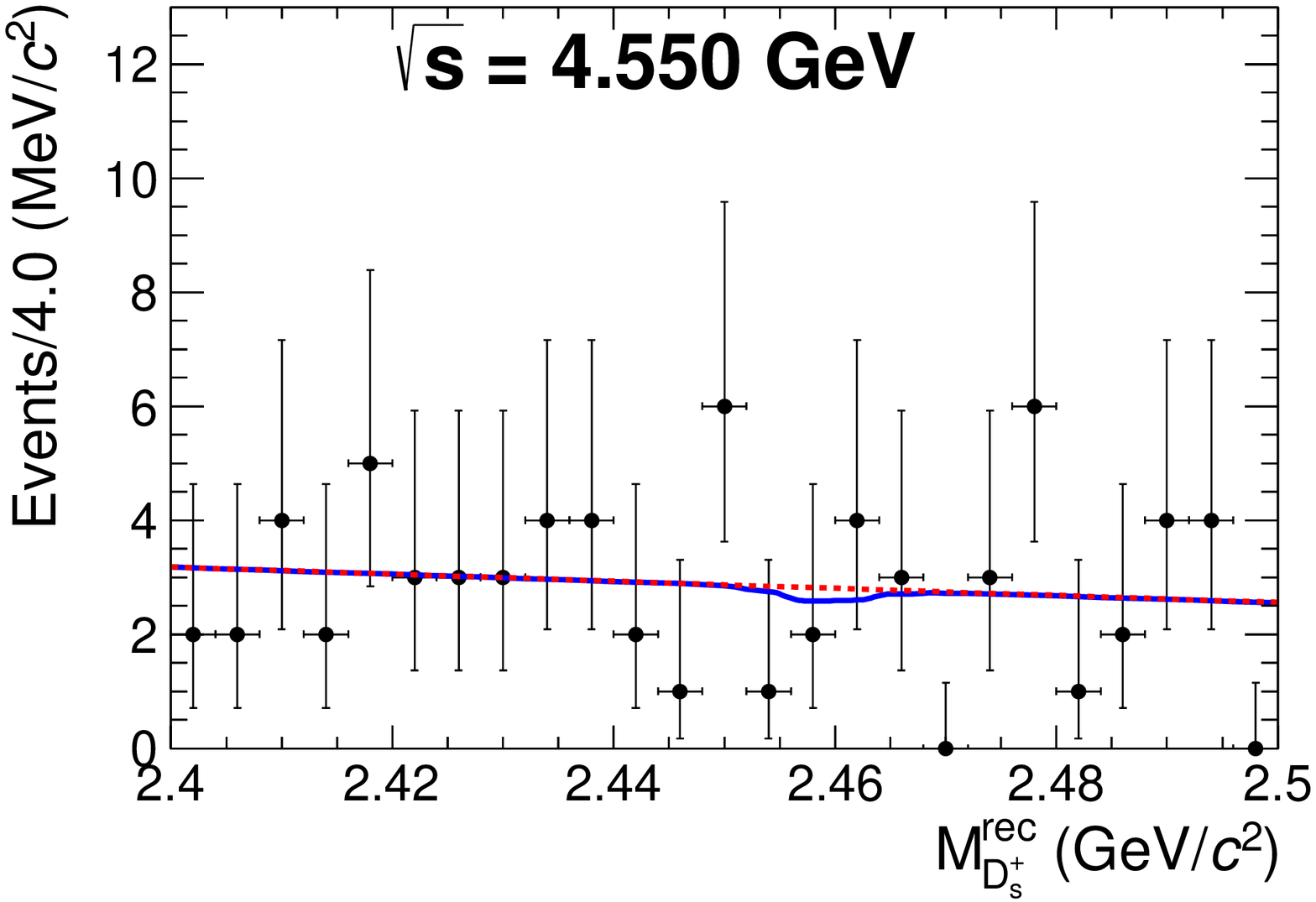}
	\includegraphics[width=.28\textwidth]{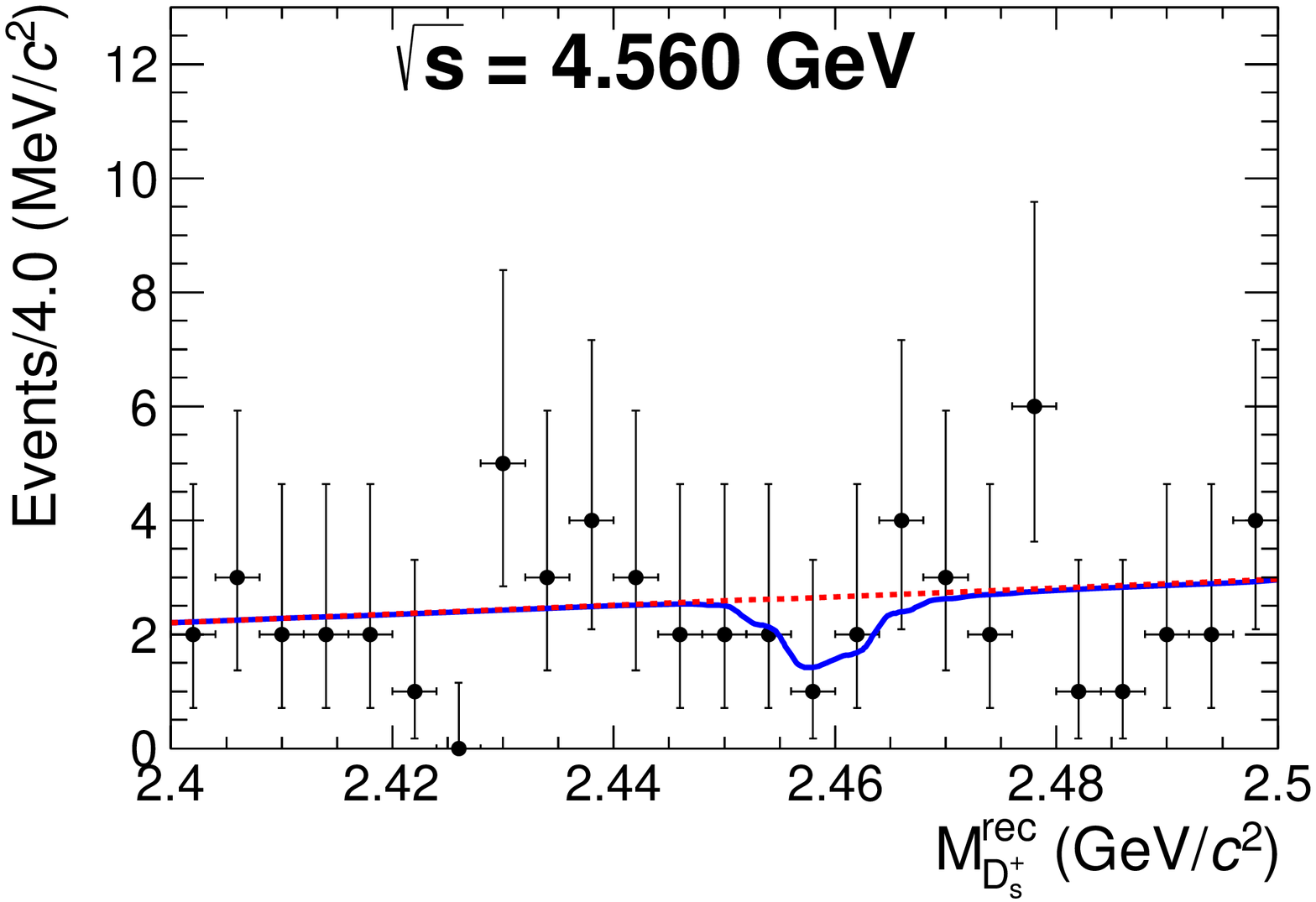}
	\includegraphics[width=.28\textwidth]{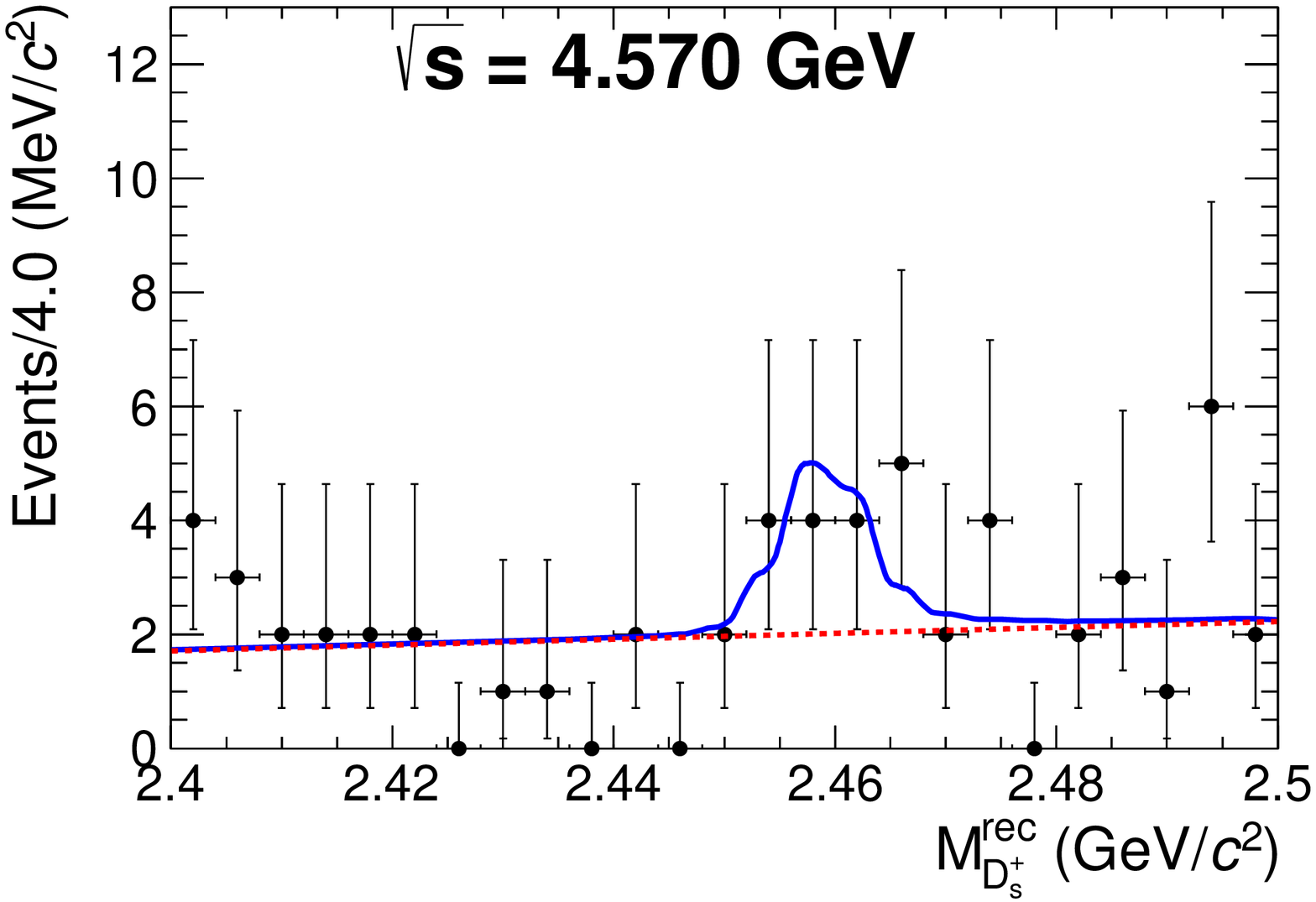}
	\includegraphics[width=.28\textwidth]{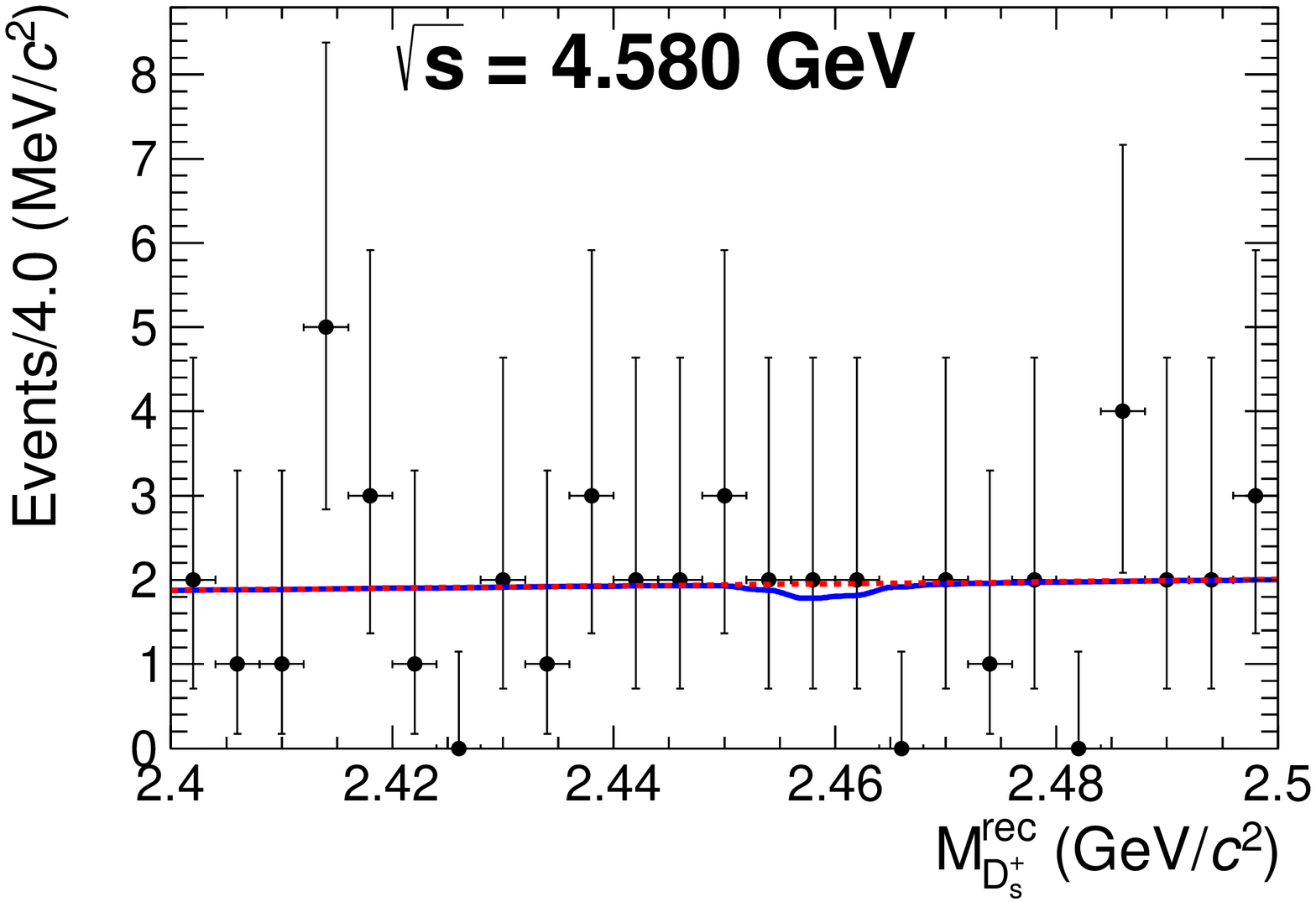}
	\includegraphics[width=.28\textwidth]{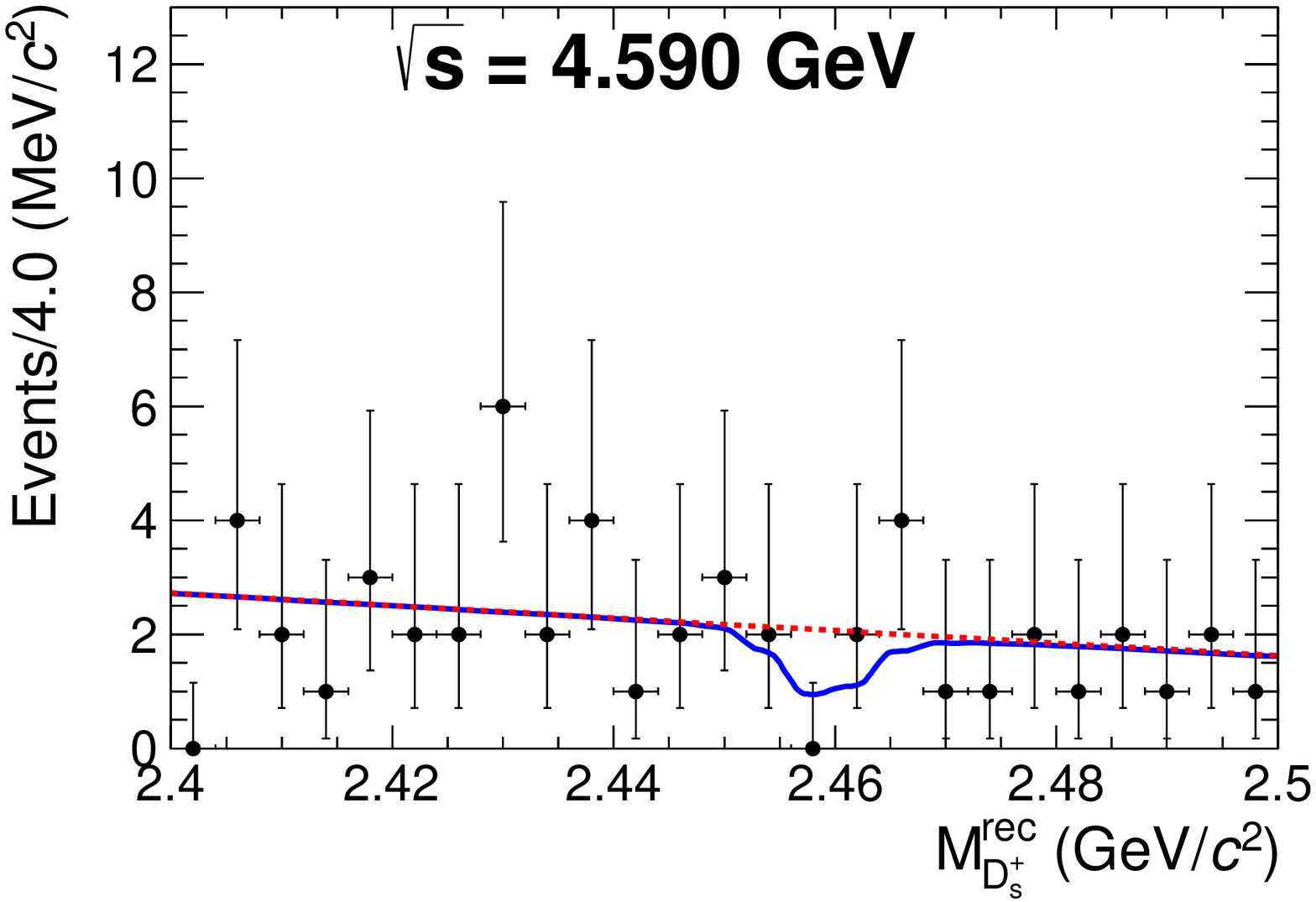}
	\caption{$M_{\Ds+}^{\rm rec}$ distributions from data samples at $\sqrt{s}= 4.550$\,GeV, 4.560\,GeV, 4.570\,GeV, 4.580\,GeV, and 4.590\,GeV. The dots with error bars are data, the solid lines are the best fits, and the dashed lines are the fitted backgrounds. The fitted results together with the signal significances are summarized in Table~\ref{tab:1}.}
	\label{fig:ds_sigma_rscan_data}
	\end{center}
\end{figure}

Due to the limited statistics, we employ a different strategy for the data samples at $\sqrt{s}=$ 4.550 GeV, 4.560 GeV, 4.570 GeV, 4.580 GeV, and 4.590\,GeV. In those cases, $M(K^+K^-\pi^+)$ is first required to satisfy $|M(K^+K^-\pi^+)-m_{D_s^+}|<10$ MeV/$c^2$. A fit is then directly performed to the $M_{\Ds+}^{\rm rec}$ distributions, using a MC-derived \Ds1- signal shape for the signal and a first-order polynomial for the background. The fit results are shown in Fig.~\ref{fig:ds_sigma_rscan_data}. No significant \Ds1- signals are observed in these five data samples. The fit results together with the signal significances are summarized in Table~\ref{tab:1}.

Since the statistical significances of the \Ds1- signal at some energy points are less than $3\sigma$, the upper limits on the numbers of \Ds1- signal events ($N_{\rm U.L.}$) are determined at the $90\%$ confidence level (C.L.) by solving the following equation:

\begin{equation}
\label{eq:UL}
      \int^{N_{\rm U.L.}}_0 \mathcal{L}(x) \, dx  \,=\,
 0.9  \int^{+\infty}_0\mathcal{L}(x) \, dx,
\end{equation}

\noindent where $x$ is the assumed yield of \Ds1- signal, and $\mathcal{L}(x)$ is the corresponding maximum likelihood from the data. The resulting $N_{\rm U.L.}$ obtained using the above method are listed in Table~\ref{tab:1}.

The Born cross section of $\ee\to\Ds+\Ds1-$ is calculated using the formula:
\begin{equation}
\begin{aligned}
\label{sigma-B}
&\sigma_{B}(\ee\to\Ds+\Ds1-)=\frac{N_{\rm fit}}{\mathcal{L}_{\rm int}(1+\delta)(1+\delta^{\rm vp})\epsilon_{D_s}},
\end{aligned}
\end{equation}

\noindent where $N_{\rm fit}$ is the \Ds1- signal yield, $1+\delta$ is the radiative correction factor obtained from a QED calculation with $1\%$ accuracy~\cite{radiator} using the {\sc kkmc} generator, $1+\delta^{\rm vp}$ is the vacuum polarization factor, whose calculations are from Ref.~\cite{vacuum} ($\delta^{\rm vp}=0.055$ for all studied energy points), and $\mathcal{L}_{\rm int}$ is the integrated luminosity at each energy point.  The product of the $D_s$ efficiency and branching fraction is
$\epsilon_{D_s}=\epsilon\mathcal{B}(D_s^+\to K^+K^-\pi^+)$
where $\epsilon$ is the detection efficiency and $\mathcal{B}(D_s^+\to K^+K^-\pi^+)$ is the branching fraction
for $D_s^+\to K^+K^-\pi^+$~\cite{PDG2018}. The calculation of the upper limits for Born cross sections at the 90\% C.L. is performed analogously, replacing $N_{\rm fit}$ with $N_{\rm U.L.}$.

\begin{figure}[htbp!]
	\begin{center}
	\includegraphics[width=.5\textwidth]{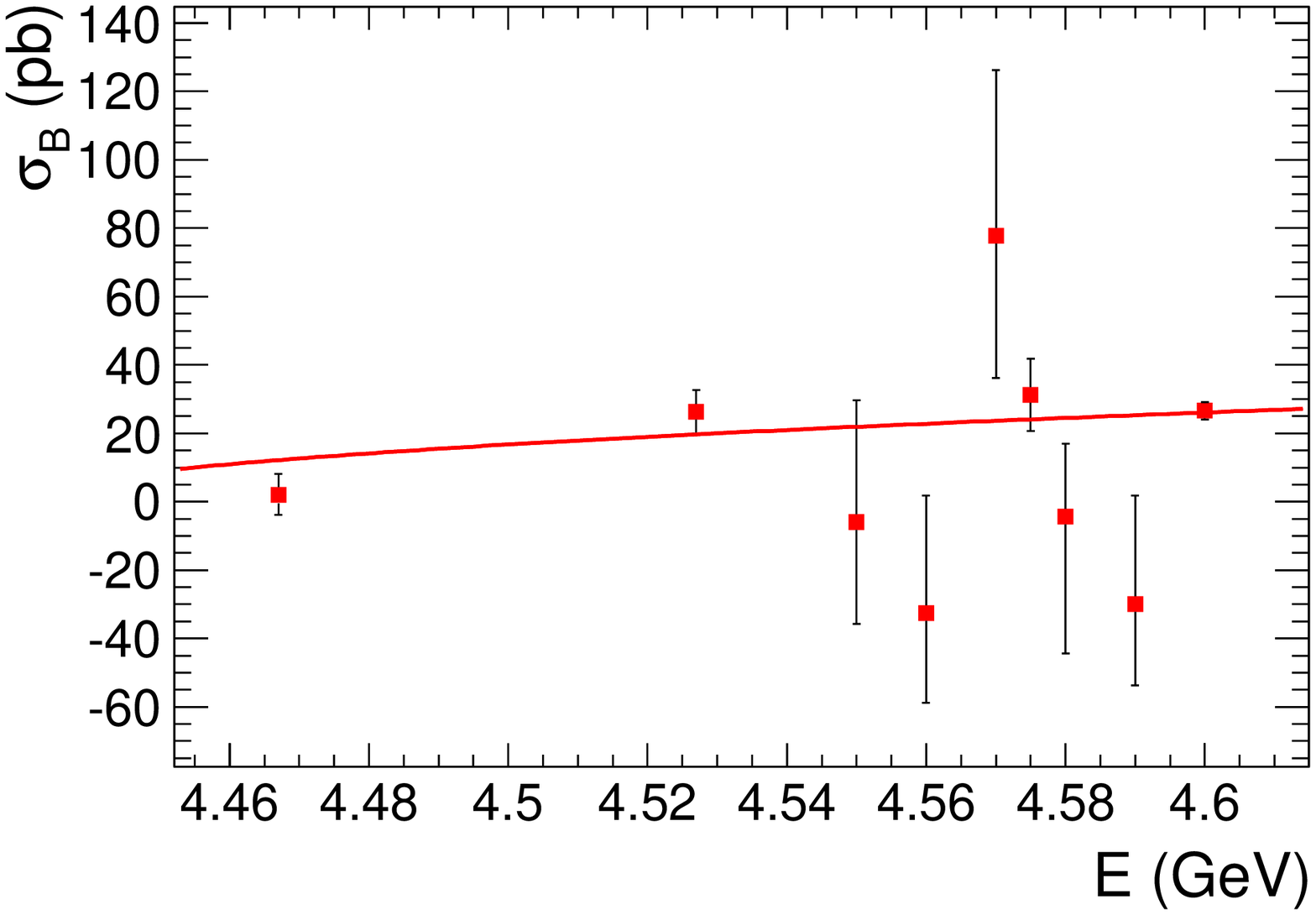}
	\caption{The fit to the Born cross sections of $\ee\to\Ds+\Ds1-$ with $\sigma\left[\ee\to\Ds\Ds1\right]\propto\sqrt{E_{\rm c.m.}-E_0}$. All error bars are statistical only.}
	\label{fig:sigma_fit}
	\end{center}
\end{figure}

The measured Born cross sections of $\ee\to\Ds+\Ds1-$ and the corresponding upper limits at the 90\% C.L.~(with systematic uncertainties included) for the energy points with signal significances less than $3\sigma$ are summarized in Table~\ref{tab:1}. The systematic uncertainties and the method to take them into account in the upper limits are discussed in Sec.~\ref{sec:sys}.
The Born cross sections with statistical error bars only are shown in Fig.~\ref{fig:sigma_fit}, together with the fit result using the prediction of Ref.~\cite{DsDs1theory}, i.e. $\sigma\left[\ee\to\Ds\Ds1\right]\propto\sqrt{E_{\rm c.m.}-E_0}$. The fit gives $\chi^2/ndf=1.75$, where $ndf$ is the number of degrees of freedom.

\section{\boldmath MEASUREMENT OF $\ee\to\Ds*+\Ds1-$ }
\begin{figure}[htbp!]
	\begin{center}
	\includegraphics[width=.4\textwidth]{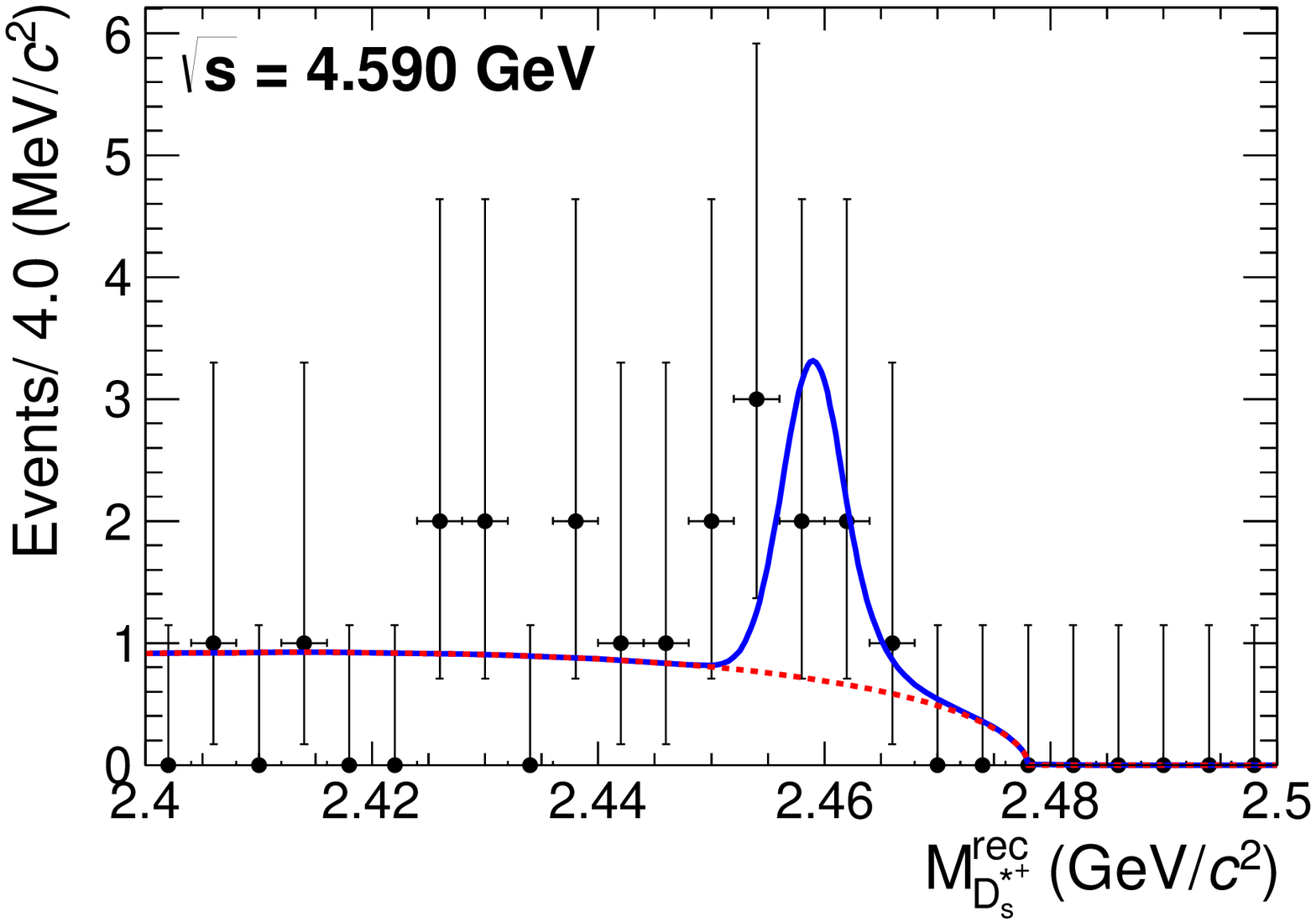}
	\includegraphics[width=.4\textwidth]{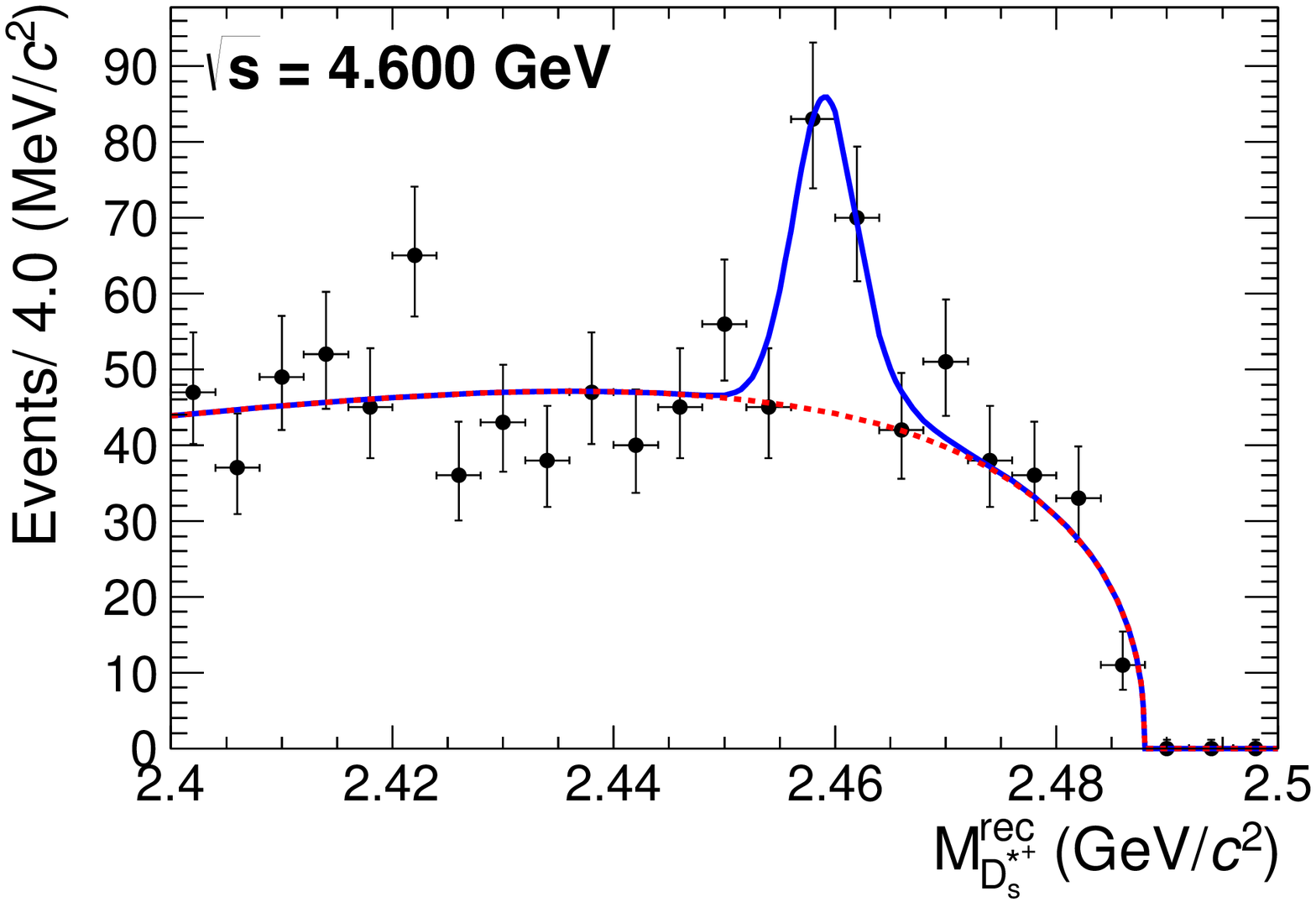}
	\caption{The $M_{\Ds*+}^{\rm rec}$ distributions from data samples at $\rts= 4.590$~GeV and 4.600~GeV, respectively;
a clear \Ds1- signal is seen at $\rts= 4.600$~\GeV.
The dots with error bars are data,
the solid line represents the best fit, and the dashed line represents the fitted background.}
	\label{fig:dss_fit}
	\end{center}
\end{figure}

In the events passing the selection criteria described in Sec.~\ref{selection}, we search for \Ds1- in the recoil mass of \Ds*+. To improve the mass resolution, mass-constrained fits to the nominal masses of \Ds+ and \Ds*+ (2C) are applied. The $\chi^2_{2C}$ is required to be less than 10 to suppress background contributions.
The recoil mass distributions of \Ds*+ from data samples at $\rts=$ 4.590~GeV and 4.600~GeV are shown in Fig.~\ref{fig:dss_fit}.
A clear \Ds1- peak is observed at $\rts=$ 4.600~GeV, while there is no clear \Ds1- signal at $\rts=$ 4.590~GeV.
Detailed study of the generic MC samples indicates that there are no peaking
background contributions in the $\Ds1-$ signal region~\cite{topo}. The background events are from the processes with $D^+ D^{*-}$, $D^0 \bar{D}^{*0}$, $\pi^0 D^+ D^{*-}$,
$\pi^- D^{*+} \bar{D}^0$, etc., in the final states.

An unbinned maximum likelihood fit is performed to the $M_{\Ds*+}^{\rm rec}$ distribution in Fig.~\ref{fig:dss_fit}. The signal is described by a Crystal Ball function~\cite{crystalball}, the parameters of which are fixed to those obtained from the fit to the $M_{\Ds*+}^{\rm rec}$ distribution in the PHSP MC sample.
The background is modeled with an ARGUS function~\cite{argus}.
The significances of the \Ds1- signal at $\rts=4.590$~GeV and 4.600~GeV are  $2.0\sigma$ and $5.9\sigma$, respectively.
The fit results together with the signal significances are summarized in Table~\ref{tab:1}.
The upper limit on the number of \Ds1- signal events $N_{\rm U.L.}$ for $\rts=$ 4.590~GeV determined at the $90\%$ C.L.\ is listed in Table~\ref{tab:1}.

The Born cross section of $\ee\to\Ds*+\Ds1-$ is calculated using the formula
\begin{equation}
\begin{aligned}
	\label{eq:dss_born}
&\sigma_{B}(\ee\to\Ds*+\Ds1-)=\frac{N_{\rm fit}}{\mathcal{L}_{\rm int}(1+\delta)(1+\delta^{\rm vp})\epsilon_{\Ds*}}.
\end{aligned}
\end{equation}

\noindent Here, the parameters have the same meaning as in Eq.~\ref{sigma-B}, except that
$\epsilon_{\Ds*}=\epsilon^* \mathcal{B}(D_s^{*+} \to \gamma D_s^+)\mathcal{B}(D_s^+ \to K^+ K^- \pi^+)$
where $\epsilon^*$ is the detection efficiency of the \Ds*+ and $\mathcal{B}(D_s^{*+} \to \gamma D_s^+)$
is the branching fraction for $D_s^{*+} \to \gamma D_s^+$~\cite{PDG2018}.

The calculated Born cross sections of $\ee\to\Ds*+\Ds1-$ at $\rts=$ 4.590 GeV and 4.600~GeV, and the upper limit at 90\% C.L.~(with systematic uncertainties included) for $\rts=$ 4.590~GeV are listed in Table~\ref{tab:1}. The systematic uncertainties are discussed in Sec.~\ref{sec:sys}.

\begin{table}[htbp!]
	\begin{center}
	\caption{Summary of the measurements of the Born cross sections for $\ee\to\Ds+\Ds1-$ and $\ee\to\Ds*+\Ds1-$. Listed in the table are the integrated luminosity $\mathcal{L}_{\rm int}$, the signal efficiency $\epsilon$ ($\epsilon^*$) from signal MC samples, the number of fitted \Ds1- signal events $N_{\rm fit}$, the 90\% C.L.\ upper limit on the number of fitted \Ds1- signal yields $N_{\rm U.L.}$, the ISR radiative correction factor $(1+\delta)$, the statistical signal significance, and the measured Born cross section $\sigma_B$ and its 90\% C.L. upper limit $\sigma_B^{\rm U.L.}$ (with systematic uncertainties included).
	}
	\label{tab:1}
	\begin{tabular}{cccD{p}{{}}{5.4}cccD{x}{{}}{4.11}}
		\hline
		$\sqrt{s}$ (GeV) & $\mathcal{L}_{\rm int}$ (pb$^{-1}$) & $\epsilon$ ($\epsilon^*$) & \multicolumn{1}{c}{$N_{\rm fit}$} & $N_{\rm U.L.}$ & $(1+\delta)$ & significance & \multicolumn{1}{c}{$\sigma_B$ ($\sigma_B^{\rm U.L.}$) (pb)} \\\hline
		\multicolumn{8}{c}{$\ee\to\Ds+\Ds1-+c.c.$} \\\hline
		4.467 & 111.1 & 32.8\% & 3.0p^{+9.8}_{-9.0} & 19.2 & 0.739 & $0.3\sigma$ & 1.9x^{+6.3}_{-5.8}\ (15.3) \\
		4.527 & 112.1 & 31.1\% & 40.0p\pm 9.7 & $\cdots$ & 0.757 & $4.9\sigma$ & 26.3x\pm 6.4\pm 2.7 \\
		4.550 & 8.8 & 30.5\% & -0.7p^{+4.2}_{-3.5} &  7.7 & 0.764 & $\cdots$ & -6.0x^{+35.7}_{-29.8}\ (67.3) \\
		4.560 & 8.3 & 30.2\% & -3.6p^{+3.8}_{-2.9} &  6.1 & 0.769 & $\cdots$ & -32.6x^{+34.4}_{-26.3}\ (62.4) \\
		4.570 & 8.4 & 30.1\% & 8.8p^{+5.5}_{-4.7} &  17.1 & 0.780 & $2.0\sigma$ & 77.7x^{+48.6}_{-41.5}\ (179) \\
		4.575 & 48.9 & 32.2\% & 22.3p\pm 7.6 & $\cdots$ & 0.788 & $3.5\sigma$ & 31.2x\pm 10.6\pm 7.0 \\
		4.580 & 8.6 & 29.9\% & -0.5p^{+2.5}_{-4.7} &  6.6 & 0.798 & $\cdots$ & -4.3x^{+21.3}_{-40.1}\ (63.9) \\
		4.590 & 8.2 & 29.6\% & -3.4p^{+3.6}_{-2.7} &  5.9 & 0.819 & $\cdots$ & -29.9x^{+31.7}_{-23.8}\ (64.2) \\
		4.600 & 586.9 & 31.8\% & 242.0p\pm 22.9 & $\cdots$ & 0.847 & $13.7\sigma$ & 26.6x\pm 2.5\pm 2.5 \\\hline
		\multicolumn{8}{c}{$\ee\to\Ds*+\Ds1-+c.c.$} \\\hline
		4.590 & 8.2 & (13.0\%) & 4.8p^{+4.8}_{-2.7} & 9.9 & 0.818 & $2.0\sigma$ & 96.7x^{+97.3}_{-54.7}\ (203) \\
		4.600 & 586.9 & (13.1\%) & 82.1p\pm 15.9 & $\cdots$ & 0.847 & $5.9\sigma$ & 22.1x\pm 4.3\pm 1.9 \\\hline
	\end{tabular}
	\end{center}
\end{table}

\section{\boldmath SYSTEMATIC UNCERTAINTIES}
\label{sec:sys}

The systematic uncertainties on the measured cross sections of $\ee\to\Ds+\Ds1-$ and $\ee\to\Ds*+\Ds1-$ come from tracking and PID efficiencies, photon detection efficiency, and MC statistics. We also consider the uncertainties from ISR and vacuum polarization corrections, the luminosity measurement, branching fractions of intermediate states, the kinematic fit, MC generator, \Ds+ mass resolution, $M_{\Ds+}^{\rm rec}$ bin width, \Ds1- mass, the background shape, and the fit range. These contributions to the systematic uncertainty are divided below into two categories:
multiplicative systematic uncertainties and additive systematic uncertainties.

Multiplicative systematic uncertainties are analyzed as follows. The uncertainties of tracking and PID are determined to be 1.5\%, 1.0\%, and 1.0\% for $K^+$, $K^-$, and $\pi^+$, respectively, using the control samples of $J/\psi\to p\bar{p}\pi^+\pi^-$ and $J/\psi\to K_S^0K^+\pi^-$, where the transverse momentum and angular region
of the signal channels are taken into account. The uncertainty of the photon reconstruction efficiency is 1.0\% per photon, which is derived from the study of $J/\psi\to\rho^0(\to \pi^+\pi^-)\pi^0(\to \gamma \gamma)$~\cite{photonpi0}. The uncertainties due to MC statistics are determined to be 1.1\% at each energy point.
The shapes of the cross section of the processes $\ee\to\Ds+\Ds1-$ and $\ee\to\Ds*+\Ds1-$ affect the radiative correction factor and the detection efficiency.
Due to the small number of data points with low statistics, a detailed determination of the energy dependence (``line shape''), which would allow for an iterative determination of radiative correction factors, is not possible.
Therefore, we change the input line shapes to a simple polynomial form, and the differences in $\varepsilon (1+\delta)$ are taken as the systematic uncertainties. The uncertainty from the vacuum polarization factor is less than 0.1\%~\cite{vacuum}, which is negligible compared to other sources of uncertainties.
The integrated luminosities of the data samples are measured using
large angle Bhabha scattering events with an uncertainty less than 1.0\%.
The uncertainties of $\mathcal{B}(D_s^+\to K^+ K^- \pi^+)$ and $\mathcal{B}(\Ds*+\to\gamma\Ds+)$ are 3.2\% and 0.7\%, respectively~\cite{PDG2018}. The uncertainty of the 2C kinematic fit is estimated using the control samples of $\ee\to\Ds*+\Ds*-$ at $\rts=$ 4.420 GeV and 4.600 GeV. The difference in the data and MC efficiencies due to the addition of the 2C kinematic fit requirement is $1.7\%$, which is taken as the systematic uncertainty.
Signal MC samples are generated with a PHSP model. We also generate signal MC samples
with a polar angle distribution of $1+\cos^2\theta$ or $1-\cos^2\theta$ for the $\Ds+/\Ds*+$
meson. The maximum differences in detection efficiencies are 1.3\% and 1.7\% for the
reconstructed $\Ds+$ and $\Ds*+$ candidates.

Additive systematic uncertainties due to the fit are analyzed as follows. The uncertainty due to the \Ds+ mass resolution is estimated by varying this mass
resolution by $\pm1\sigma$ when fitting the $K^+K^-\pi^+$ invariant mass distributions in $M_{\Ds+}^{\rm rec}$ bins. The differences in the fitted \Ds1- signal yields
are taken as the systematic uncertainties. The uncertainties due to the $M_{\Ds+}^{\rm rec}$ bin width are studied by varying the $M_{\Ds+}^{\rm rec}$ bin width from $4.0~\MeVcc$ to $5.0~\MeVcc$. The differences in the fitted \Ds1- signal yields are taken as the systematic uncertainties.
The uncertainties due to the \Ds1- mass are obtained by varying the  \Ds1- mass by $\pm1\sigma$, i.e.~$0.6~\MeVcc$~\cite{PDG2018},
in the fit of the $M_{\Ds+}^{\rm rec}$ distribution.
The difference in the fitted \Ds1- signal yields is taken as the systematic uncertainty.
In the analysis of $\ee\to\Ds+\Ds1-$, the uncertainties attributed to the background shape
are estimated by using different background shapes: (1) a first-order polynomial is used as the background shape
(for $\sqrt{s}=$ 4.527 GeV and 4.600 GeV data samples, a third-order polynomial is used as the background shape); (2) a second-order polynomial and the normalized contribution from $\ee\to\Ds*+\Ds*-$ are used as the total background shape. In the analysis of $\ee\to\Ds*+\Ds1-$, the uncertainties due to the background shape are estimated by using a parameterized polynomial $f(M)=(M-M_a)^c(M_b-M)^d$ instead of an ARGUS function~\cite{argus}, where $M_a$ and $M_b$ are the lower and upper thresholds of the \Ds*+ recoil mass distribution. The maximum differences in the fitted \Ds1- signal yields are considered as the systematic uncertainties. In the analysis of $\ee\to\Ds+\Ds1-$, the uncertainties due to the fit range are obtained by varying the fit range by 10 MeV on the left or right side. In the analysis of $\ee\to\Ds*+\Ds1-$, the uncertainties due to the fit range are determined by varying the fit range from [2.40, 2.49]~\GeVcc\ to [2.30, 2.49]~\GeVcc. The differences in the fitted \Ds1- signal yields are taken as the systematic uncertainties.

For those energy points with a statistical significance larger than $3\sigma$,
the central values of the cross section with statistical and systematic uncertainties are reported,
and all of the systematic uncertainties are summarized in Table~\ref{tab:sys1}. For the other energy points with \Ds1- signal significance less than $3\sigma$,
the upper limits  on the cross section at the 90\% C.L. are reported and the systematic uncertainties are
taken into account in two steps. First, when we study the additive systematic uncertainties described above,
we take the most conservative upper limit at the 90\% C.L. on the number of \Ds1- signal yields.
Then, to take into account the multiplicative systematic uncertainty,
the likelihood with the most conservative upper limit is convolved with a Gaussian function, with a width
equal to the corresponding total multiplicative systematic uncertainty.
All of the multiplicative systematic uncertainties for the energy points with \Ds1- signal significance less than $3\sigma$ are summarized in Table~\ref{tab:sys2}. Assuming that all the sources are
independent, the total systematic uncertainty is obtained by adding them in quadrature.
The final results of the Born cross section with systematic uncertainties considered are listed in Table~\ref{tab:1}. The comparison of the Born cross sections of $\ee\to\Ds+\Ds1-$ and $\ee\to\Ds*+\Ds1-$ is shown in Fig.~\ref{fig:ds_sigma} with statistical error bars only.

\begin{table}[htbp!]
	\centering
	\caption{Summary of systematic uncertainties of the Born cross sections of $\ee\to\Ds+\Ds1-$ and $\ee\to\Ds*+\Ds1-$ for those energy points
with statistical significances larger than 3$\sigma$.}
	\label{tab:sys1}
	\begin{tabular}{cD{\%}{\%}{4.1}D{\%}{\%}{4.1}D{\%}{\%}{3.1}D{\%}{\%}{3.1}}
		\hline
		Sources & \multicolumn{3}{c}{$\ee\to\Ds+\Ds1-$} & \multicolumn{1}{c}{$\ee\to\Ds*+\Ds1-$} \\\hline
		$\rts$ (GeV) & \multicolumn{1}{c}{4.527} & \multicolumn{1}{c}{4.575} & \multicolumn{1}{c}{4.600} & \multicolumn{1}{c}{4.600} \\\hline
		Tracking, PID and photon & 3.5\% & 3.5\% & 3.5\% & 3.7\% \\
		MC statistics & 0.5\% & 0.5\% & 0.5\% & 1.0\% \\
		ISR correction & 4.6\% & 8.2\% & 5.5\% & 0.1\% \\
		Luminosity & 0.7\% & 0.7\% & 0.7\% & 0.7\% \\
		Branching fraction & 3.2\% & 3.2\% & 3.2\% & 3.3\% \\
		Kinematic fit & \multicolumn{1}{c}{$\cdots$} & \multicolumn{1}{c}{$\cdots$} & \multicolumn{1}{c}{$\cdots$} & 1.7\% \\
        MC generator & 1.3\% & 1.3\% & 1.3\% & 1.7\% \\
		\Ds+ mass resolution & 1.3\% & 4.4\% & 1.5\% & \multicolumn{1}{c}{$\cdots$} \\
		$M_{\Ds+}^{\rm rec}$ bin width & 6.1\% & 13.6\% & 1.5\% & \multicolumn{1}{c}{$\cdots$} \\
		\Ds1- mass & 0.9\% & 11.8\% & 2.6\% & \multicolumn{1}{c}{$\cdots$} \\
		Background shape & 2.9\% & 5.5\% & 4.1\% & 1.7\% \\
		Fit range & 2.5\% & 5.6\% & 1.1\% & 5.9\% \\\hline
		Total & 10.1\% & 22.3\% & 9.2\% & 8.3\% \\\hline
	\end{tabular}
\end{table}

\begin{table}[htbp!]
	\centering
	\caption{Summary of multiplicative systematic uncertainties of the Born cross sections of $\ee\to\Ds+\Ds1-$ and $\ee\to\Ds*+\Ds1-$ for those energy points with statistical significances less than 3$\sigma$.}
	\label{tab:sys2}
	\begin{tabular}{cD{\%}{\%}{4.1}D{\%}{\%}{3.1}D{\%}{\%}{3.1}D{\%}{\%}{3.1}D{\%}{\%}{3.1}D{\%}{\%}{3.1}D{\%}{\%}{3.1}}
		\hline
		Sources & \multicolumn{6}{c}{$\ee\to\Ds+\Ds1-$} & \multicolumn{1}{c}{$\ee\to\Ds*+\Ds1-$} \\\hline
		$\rts$ (GeV) & \multicolumn{1}{c}{4.467} & \multicolumn{1}{c}{4.550} & \multicolumn{1}{c}{4.560} & \multicolumn{1}{c}{4.570} & \multicolumn{1}{c}{4.580} & \multicolumn{1}{c}{4.590} & \multicolumn{1}{c}{4.590} \\\hline
		Tracking, PID and photon & 3.5\% & 3.5\% & 3.5\% & 3.5\% & 3.5\% & 3.5\% & 3.7\% \\
		MC statistics & 0.5\% & 0.5\% & 0.5\% & 0.5\% & 0.5\% & 0.5\% & 1.1\% \\
		ISR correction & 13.1\% & 7.6\% & 8.1\% & 2.8\% & 7.6\% & 7.0\% & 1.6\% \\
		Luminosity & 0.7\% & 0.8\% & 0.8\% & 0.8\% & 0.7\% & 0.7\% & 0.7\% \\
		Branching fraction & 3.2\% & 3.2\% & 3.2\% & 3.2\% & 3.2\% & 3.2\% & 3.3\% \\
		Kinematic fit & \multicolumn{1}{c}{$\cdots$} & \multicolumn{1}{c}{$\cdots$} & \multicolumn{1}{c}{$\cdots$} & \multicolumn{1}{c}{$\cdots$} & \multicolumn{1}{c}{$\cdots$} & \multicolumn{1}{c}{$\cdots$} & 1.7\% \\
MC generator & 1.3\% & 1.3\% & 1.3\% & 1.3\% & 1.3\% & 1.3\% & 1.7\% \\ \hline
		Total & 14.0\% & 9.1\% & 9.5\% & 5.7\% & 9.1\% & 8.6\% & 5.7\% \\\hline
	\end{tabular}
\end{table}

\section{\boldmath SUMMARY}

\begin{figure}[htbp!]
	\begin{center}
	\includegraphics[width=.7\textwidth]{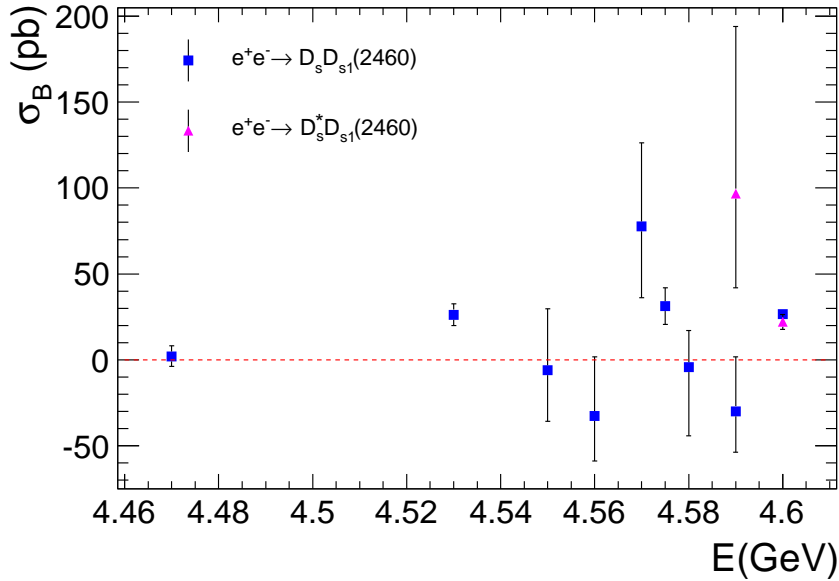}
	\caption{The comparison of the Born cross sections of $\ee\to\Ds+\Ds1-$ (squares with error bars) and $\ee\to\Ds*+\Ds1-$ (triangles with error bars), where the error bars are statistical only.}
	\label{fig:ds_sigma}
	\end{center}
\end{figure}

In summary, we observe \Ds1- signals with statistical significances larger than $3\sigma$
in the processes $\ee\to\Ds+\Ds1-$ ($\ee\to\Ds*+\Ds1-$) at c.m.\ energies of 4.527 GeV,
4.575 GeV, and 4.600 GeV (4.600 GeV). The Born cross sections, $\sigma_B[\ee\to\Ds+\Ds1-]$ and $\sigma_B[\ee\to\Ds*+\Ds1-]$,
have been measured for the first time and displayed in Fig.~\ref{fig:ds_sigma}.
The prediction on the energy dependence of the Born cross section given in Ref.~\cite{DsDs1theory},
i.e.~$\sigma\left[\ee\to\Ds\Ds1\right]\propto\sqrt{E_{\rm c.m.}-E_0}$, is
confronted with the result of our measurement in Fig.~\ref{fig:sigma_fit}.
Within the statistical uncertainty of the measurement, the theoretical prediction can describe the data.

\begin{acknowledgments}
The BESIII collaboration thanks the staff of BEPCII and the IHEP computing center for their strong support. This work is supported in part by National Key Basic Research Program of China under Contract No. 2015CB856700; National Natural Science Foundation of China (NSFC) under Contracts Nos. 11625523, 11635010, 11735014, 11822506, 11835012, 11935015, 11935016, 11935018, 11961141012; the Chinese Academy of Sciences (CAS) Large-Scale Scientific Facility Program; Joint Large-Scale Scientific Facility Funds of the NSFC and CAS under Contracts Nos. U1732263, U1832207; CAS Key Research Program of Frontier Sciences under Contracts Nos. QYZDJ-SSW-SLH003, QYZDJ-SSW-SLH040; 100 Talents Program of CAS; INPAC and Shanghai Key Laboratory for Particle Physics and Cosmology; ERC under Contract No. 758462; German Research Foundation DFG under Contracts Nos. Collaborative Research Center CRC 1044, FOR 2359; Istituto Nazionale di Fisica Nucleare, Italy; Ministry of Development of Turkey under Contract No. DPT2006K-120470; National Science and Technology fund; STFC (United Kingdom); The Knut and Alice Wallenberg Foundation (Sweden) under Contract No. 2016.0157; The Royal Society, UK under Contracts Nos. DH140054, DH160214; The Swedish Research Council; U. S. Department of Energy under Contracts Nos. DE-FG02-05ER41374, DE-SC-0012069.
\end{acknowledgments}

\end{document}